\documentclass[aps,prl,twocolumn,showpacs,preprintnumbers,amsmath,amssymb,superscriptaddress,floats,nofootinbib]{revtex4-1}
\usepackage{graphicx,color}
\begin{document}

\def\Journal#1#2#3#4{{#1} {\bf #2}, #3 (#4)}
\def\NCA{\rm Nuovo Cimento}
\def\NPA{{\rm Nucl. Phys.} A}
\def\NIM{\rm Nucl. Instrum. Methods}
\def\NIMA{{\rm Nucl. Instrum. Methods} A}
\def\NPB{{\rm Nucl. Phys.} B}
\def\PLB{{\rm Phys. Lett.}  B}
\def\PRL{\rm Phys. Rev. Lett.}
\def\PRD{{\rm Phys. Rev.} D}
\def\PRC{{\rm Phys. Rev.} C}
\def\ZPC{{\rm Z. Phys.} C}
\def\JPG{{\rm J. Phys.} G}


\title{Unitarity Tests of the Neutrino Mixing Matrix }
\date{\today}
\author{X. Qian}\email[Corresponding author: ]{xqian@bnl.gov}
\author{C. Zhang}
\author{M. Diwan}
\affiliation{Physics Department, Brookhaven National Laboratory, Upton, NY}
\author{P. Vogel}
\affiliation{Kellogg Radiation Laboratory, California Institute of Technology, Pasadena, CA}

\begin{abstract}
We discuss unitarity tests of the neutrino mixing (PMNS) matrix.  
We show that the combination of solar neutrino experiments, medium-baseline and short-baseline reactor 
antineutrino experiments make it possible to perform the first direct unitarity test of the PMNS matrix.
In particular, the measurements of Daya Bay and JUNO (a next generation 
medium-baseline reactor experiment) will lay the foundation of a precise unitarity 
test of $|U_{e1}|^2 + |U_{e2}|^2 + |U_{e3}|^2 = 1 $. Furthermore, the precision measurement of 
$\sin^22\theta_{13}$ in both the $\bar{\nu}_e$ disappearance and the $\nu_e$ appearance 
(from a $\nu_{\mu}$ beam) channels will provide an indirect unitarity test of the PMNS matrix. 
Together with the search for appearance/disappearance at very short distances, these tests could 
provide important information about the possible new physics beyond the three-neutrino model.
\end{abstract}
\maketitle

\thispagestyle{plain}
\paragraph{\bf Introduction:} 
In the past decades our understanding of neutrinos has  advanced dramatically. 
Initially, neutrinos were thought to be massless, since only left-handed 
neutrinos and right-handed antineutrinos were detected 
in experiments~\cite{helicity}. The existence of non-zero neutrino masses 
and the neutrino mixing were then successfully established
through the observation of neutrino flavor oscillations. 
Recent reviews can be found e.g. in Ref.~\cite{PDG,mckeown_review}. 
In the three-neutrino framework, the oscillations 
are characterized by the neutrino mixing (commonly referred to as the 
Pontecorvo-Maki-Nakagawa-Sakata or PMNS in short) 
matrix~\cite{ponte1,ponte2,Maki} and two neutrino mass-squared 
differences ($\Delta m^2_{32} = m^2_{3}-m^2_{2}$ and 
$\Delta m^2_{21} = m^2_{2}-m^2_{1}$). 

The PMNS matrix $U_{PMNS}$ (or $U$ in short),
\begin{equation}
\left ( 
\begin{matrix}
\nu_{e} \\
\nu_{\mu} \\
\nu_{\tau}
\end{matrix}
\right) = \left( 
\begin{matrix}
U_{e1} & U_{e2} & U_{e3} \\
U_{\mu 1} & U_{\mu 2} & U_{\mu 3} \\
U_{\tau 1} & U_{\tau 2} & U_{\tau 3} 
\end{matrix}
\right)
\cdot \left(
\begin{matrix}
\nu_1 \\
\nu_2 \\
\nu_3
\end{matrix}
\right),
\end{equation}
describes the mixing between the neutrino flavor ($\nu_e$, $\nu_\mu$, $\nu_\tau$) and mass eigenstates 
($\nu_1$, $\nu_2$, and $\nu_3$ with masses $m_1$, $m_2$, and $m_3$, respectively).  Components of the 
PMNS matrix can be determined through measurements of neutrino oscillations. For neutrinos 
with energy $E$ and flavor $l$, the probability of its transformation to flavor $l'$ after traveling a 
distance $L$ in vacuum is expressed as:
\begin{eqnarray}\label{eq:osc_dis}
P(\nu_l\rightarrow \nu_{l'}) = \left |\sum_{i} U_{li}U^{*}_{l'i}e^{-i(m_{i}^2/2E)L} \right | ^2  \nonumber \\
= \sum_{i}|U_{li}U^*_{l'i}|^2 + \Re \sum_{i} \sum_{j \neq i} U_{li} U^{*}_{l'i} U^{*}_{lj} U_{l'j} e^{i\frac{\Delta m^2_{ij} L}{2E}}.
\end{eqnarray}

The unitarity tests of the PMNS matrix refer to establishing whether $U \times U^* \stackrel{?}{=} I$ and $U^* \times U \stackrel{?}{=} I$, where $I$ is the 
3$\times$3 unit matrix. These conditions  are represented by twelve equations in total:
\begin{eqnarray}
|U_{l1}|^2 + |U_{l2}|^2 + |U_{l3}|^2 &\stackrel{?}{=}& 1 |_{l=e,\mu,\tau} \label{eq:uni1}\\
U_{l1}U^{*}_{l'1} + U_{l2}U^{*}_{l'2} + U_{l3}U^{*}_{l'3} &\stackrel{?}{=}& 0 |_{l,l'=e,\mu,\tau; l'\neq l} \label{eq:uni2}\\
|U_{e i}|^2 + |U_{\mu i}|^2 + |U_{\tau i}|^2 &\stackrel{?}{=}& 1 |_{i=1,2,3} \label{eq:uni3}\\
U_{e i}U^{*}_{e j} + U_{\mu i}U^{*}_{\mu j} + U_{\tau i}U^{*}_{\tau j} &\stackrel{?}{=}& 0 |_{i,j=1,2,3;i\neq j}.\label{eq:uni4}
\end{eqnarray}

The PMNS matrix is conventionally written as explicitly unitary: 
\begin{equation} 
\left( \begin{matrix}
c_{12}c_{13} & s_{12}c_{13} & s_{13}e^{-i\delta} \\
-s_{12}c_{23}-c_{12}s_{23}s_{13}e^{i\delta} & c_{12}c_{23}-s_{12}s_{23}s_{13}e^{i\delta} & s_{23}c_{13} \\
s_{12}s_{23}-c_{12}c_{23}s_{13}e^{i\delta} & -c_{12}s_{23}-s_{12}c_{23}s_{13}e^{i\delta} & c_{23}c_{13}
\end{matrix}
\right),  \nonumber
\end{equation}
with three mixing angles  $\theta_{12}$, $\theta_{23}$, $\theta_{13}$ 
($s_{ij} = \sin\theta_{ij}$ and $c_{ij} = \cos\theta_{ij}$), and a phase $\delta$, 
commonly referred to as the CP phase in the leptonic sector.

Super-Kamiokande~\cite{SuperK}, K2K~\cite{K2K}, MINOS~\cite{MINOS},  T2K~\cite{T2K_muon}, and
IceCube~\cite{icecube} experiments 
determined the angle $\theta_{23}$ and the mass difference $|\Delta m^2_{32}|$ using the $\nu_\mu$ disappearance channel  
with atmospheric and accelerator neutrinos. The KamLAND~\cite{KamLAND} and SNO~\cite{SNO} experiments
measured $\theta_{12}$ and $\Delta m^2_{21}$ with $\bar{\nu}_e$ disappearance channel using reactor 
antineutrinos and $\nu_e$ disappearance channel using solar neutrinos~\footnote{\textcolor{black}{Due to the MSW effect, the $\nu_e$
disappearance probability in SNO is different from Eq.~\eqref{eq:osc_dis}.}}, respectively. 
Recently, the Daya Bay~\cite{dayabay,dayabay_cpc}, Double Chooz~\cite{doublec}, and RENO~\cite{reno}
measured $\theta_{13}$ and are on their ways to measure $|\Delta m^2_{31}|$ with $\bar{\nu}_e$ disappearance
using reactor antineutrinos. The current knowledge of the mixing angles and mass squared 
differences~\cite{PDG} are~\footnote{The uncertainties 
represent the 68\% confidence intervals. The limit quoted for $\sin^22\theta_{23}$ corresponds
to the projection of the 90\% confidence interval in the $\sin^22\theta_{23}$-$\Delta m^2_{23}$ 
plane onto the $\sin^22\theta_{23}$ axis.}:
\begin{eqnarray}
\sin^2 2\theta_{12} &=& 0.857\pm0.024 \nonumber \\
\sin^2 2\theta_{23} &>& 0.95 \nonumber \\
\sin^2 2\theta_{13} &=& 0.095 \pm 0.010 \nonumber \\ 
\Delta m^2_{21} &=& (7.5\pm 0.2)\times 10^{-5} eV^2 \nonumber \\
|\Delta m^2_{32}| &=& (2.32^{+0.12}_{-0.08})\times 10^{-3} eV^2.
\end{eqnarray}

Besides the disappearance channels listed above, the appearance channel is becoming a powerful
tool to determine the matrix elements of $U_{PMNS}$.  The MINOS~\cite{MINOS_ap} and T2K~\cite{T2K_ap,T2K_ap1} 
measured the $\nu_\mu$ to $\nu_e$ appearance probability with accelerator neutrinos. 
In particular, T2K~\cite{T2K_ap1} established the $\nu_e$ appearance 
at a 7.5$\sigma$ level. OPERA~\cite{opera} and Super-Kamiokande~\cite{SK_tau} 
experiments observed the $\nu_\mu$ to $\nu_{\tau}$ appearance with accelerator 
and atmospheric neutrinos, respectively. In the neutrino sector the determination of the
remaining unknown quantities, 
including the value of CP phase $\delta$ and the sign of $|\Delta m^2_{32}|$ 
(neutrino mass hierarchy), are the goals of the current (No$\nu$a~\cite{nova} and T2K), and next 
generation neutrino oscillation experiments (LBNE~\cite{LBNE,LBNE_off}, JUNO~\cite{JUNO1,JUNO2},
Hyper-K~\cite{hyperK}, and PINGU~\cite{PINGU}). With future precise measurements of neutrino 
oscillation characteristics, unitarity tests of the PMNS matrix become possible. In the following,
we will discuss the direct and indirect unitarity tests of the PMNS matrix. 

\paragraph{\bf Direct unitarity test of the first row:}

In a direct unitarity test, individual components of the PMNS matrix
are measured.  Eqs.~\eqref{eq:uni1}-\eqref{eq:uni4} may then be tested 
directly. The most promising one is the test of the first row 
$|U_{e1}|^2+|U_{e2}|^2+|U_{e3}|^2 \stackrel{?}{=} 1$,
to be discussed in detail below.
 
SNO measurement of $\nu_e$ disappearance with solar neutrinos provides
the first constraint.  The higher energy $^8B$ solar neutrinos 
detected in SNO can be well approximated as the mass eigenstates due to 
the MSW effect~\cite{msw1,msw2,msw3,msw4}. 
Therefore, by comparing the charged current ($\nu_e$ only) and the neutral current 
(sum of all $\nu_{l}$) events, a direct constraint on
$\cos^{4}\theta_{13}\sin^2\theta_{12} + \sin^{4}\theta_{13}$ or in terms
of the PMNS matrix elements on the combination 
$|U_{e2}|^2\cdot(|U_{e1}|^2+|U_{e2}|^2) + |U_{e3}|^4$
is provided.

In addition, $|U_{e1}|^2$, $|U_{e2}|^2$, and $|U_{e3}|^2$ can be constrained 
using the $\bar{\nu}_e$ disappearance with reactor experiments. 
In particular, $4|U_{e1}|^2\cdot|U_{e2}|^2$ was first determined by the the KamLAND 
experiment, and will be significantly improved by the next generation medium-baseline 
reactor antineutrino experiments (e.g., the upcoming "Jianmen Underground Neutrino 
Observatory" (JUNO) experiment, and the RENO-50 experiment). 
Meanwhile, the currently running Daya Bay experiment (together with RENO and Double Chooz) 
will provide the most precise measurement of the $\bar{\nu}_e$ disappearance for 
the oscillations governed by $\Delta m^2_{3x}$ ($\Delta m^2_{31}$ and $\Delta m^2_{32}$), 
constraining $4|U_{e3}|^2\cdot(|U_{e1}|^2+|U_{e2}|^2)$. However, due to the tiny difference 
between $\Delta m^2_{31}$ and $\Delta m^2_{32}$, the Daya Bay experiment can 
not determine $4|U_{e3}|^2 \cdot |U_{e1}|^2$ and $4|U_{e3}|^2 \cdot |U_{e2}|$ 
separately. With three independent constrains, three unknown $|U_{e1}|^2$, $|U_{e2}|^2$, 
and $|U_{e3}|^2$ can be completely determined. Therefore, the combination of 
medium-baseline reactor experiments (KamLAND, JUNO and RENO-50), 
short-baseline reactor experiments (Daya Bay, RENO, and Double Chooz), 
and SNO solar neutrino results make possible the first direct unitarity test 
of the PMNS matrix. 

In the following, as an example, we study the expected sensitivity of testing  
$|U_{e1}|^2+|U_{e2}|^2+|U_{e3}|^2 \stackrel{?}{=} 1$ with SNO, Daya Bay, 
and JUNO. We adapted the fitter developed in Ref.~\cite{JUNO1}, 
which is used to study the physics capability of JUNO.
The expected results from Daya Bay and the results of SNO are taken 
from Ref.~\cite{dayabay_proceeding} and Ref.~\cite{SNO}, respectively. For JUNO, 
a 20 kt fiducial volume liquid scintillator detector is assumed at a distance of 
55 km from the reactor complex with a total thermal power of 40 GW and five years running 
time. 

The experimental uncertainties in the absolute normalization in 
both the detection efficiency and the neutrino flux have to be taken into 
account for Daya Bay and JUNO. In particular,
the debate regarding the "reactor anomaly"~\cite{anom, anom1} shows that 
the uncertainty in reactor flux can be as large as 6-8\%. In Daya Bay,
the uncertainty in reactor flux is mitigated by using the ratio 
method~\cite{ratio} with near and far detectors. In JUNO, the constraint
of $4|U_{e1}|^2\cdot|U_{e2}|^2$ is mainly coming from the spectrum 
distortion~\cite{JUNO1} due to the $\Delta m^2_{21}$ oscillation. 
In both cases, the oscillation formula need to be modified 
and becomes  actually
\begin{eqnarray}\label{eq:osc1}
P &=& \left(|U_{e1}|^2+|U_{e2}|^2+|U_{e3}|^2\right)^2  \nonumber \\
&\cdot& ( 1 - \frac{4|U_{e1}|^2|U_{e2}|^2}{ (|U_{e1}|^2+|U_{e2}|^2 + |U_{e3}|^2)^2} \sin^2 \left( \frac{\Delta m^{2}_{21} L }{4E} \right) \nonumber \\
&-& \frac{4|U_{e3}|^2(|U_{e1}|^2+|U_{e2}|^2)}{(|U_{e1}|^2+|U_{e2}|^2 + |U_{e3}|^2)^2} \sin^2 \left( \frac{\Delta m^{2}_{3x} L }{4E} \right) ),
\end{eqnarray}
in which the overall $(|U_{e1}|^2+|U_{e2}|^2+|U_{e3}|^2)^2$ term cannot be separated
from the uncertainty in the absolute normalization.
Therefore, instead of 
constraining $4|U_{e3}|^2\cdot(|U_{e1}|^2+|U_{e2}|)$ and $4|U_{e1}|^2\cdot|U_{e2}|^2$,
the Daya Bay and JUNO experiments are in fact constraining and going to constrain 
$\frac{4|U_{e3}|^2\cdot(|U_{e1}|^2+|U_{e2}|^2)}{(|U_{e1}|^2+|U_{e2}|^2 + |U_{e3}|^2)^2}$ and 
$\frac{4|U_{e1}|^2\cdot|U_{e2}|^2}{(|U_{e1}|^2+|U_{e2}|^2 + |U_{e3}|^2)^2}$, respectively. 
 For example, if the 6\% reactor anomaly is due to the existence of heavy sterile neutrinos, the impact
of the fast oscillations of the sterile neutrino components will be absorbed into
$(|U_{e1}|^2+|U_{e2}|^2+|U_{e3}|^2)^2 \simeq 0.94$. The $\theta_{13}$ ($\theta_{12}$) angle 
measured by Daya Bay (JUNO) would therefore be an effective angle: $\sin^22\theta^{eff}_{13} 
= \frac{4|U_{e3}|^2\cdot \left(|U_{e1}|^2+|U_{e2}|^2 \right)}{0.94}$ 
instead of $4|U_{e3}|^2\cdot \left(|U_{e1}|^2+|U_{e2}|^2 \right)$ ($\sin^22\theta^{eff}_{12} \approx \frac{4|U_{e1}|^2|U_{e2}|^2}{0.94}$).

\begin{figure}[htb]
\centering
\includegraphics[width=90mm]{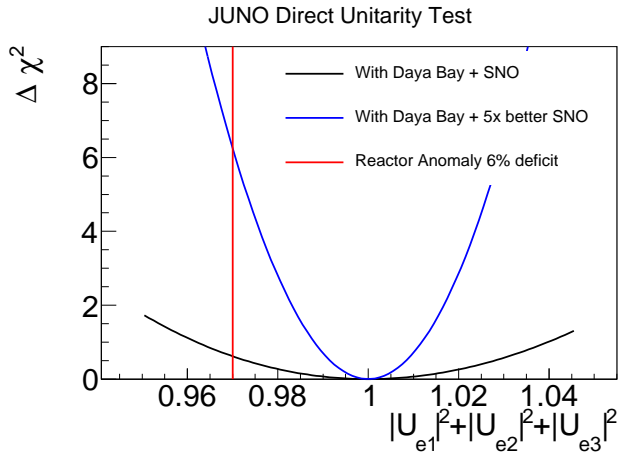}
\caption{Direct unitarity test of $|U_{e1}|^2+|U_{e2}|^2+|U_{e3}|^2 \stackrel{?}{=} 1$ by 
combining JUNO, Daya Bay, and solar results. We considered two scenarios i) current SNO 
constraint  and ii) a five times better constraint than SNO. 
In addition, the red line shows the suggested value of 
$|U_{e1}|^2+|U_{e2}|^2+|U_{e3}|^2$ given a 6\% reactor anomaly. 
See the text for more discussions.}
\label{fig:juno_direct}
\end{figure}

In Fig.~\ref{fig:juno_direct}, the sensitivity of the direct unitarity test of the
$|U_{e1}|^2+|U_{e2}|^2+|U_{e3}|^2$ are shown after combining the expected results of JUNO, 
Daya Bay, and the current SNO results.  It is assumed  that Daya Bay will reach 
 $\frac{4|U_{e3}|^2\cdot(|U_{e1}|^2+|U_{e2}|)}{(|U_{e1}|^2+|U_{e2}|^2 
+ |U_{e3}|^2)^2} = 0.09 \pm 0.0035$~\cite{dayabay_proceeding}. For the purpose of this study, we approximate 
SNO results as $|U_{e2}|^2\cdot(|U_{e1}|^2+|U_{e2}|^2) + |U_{e3}|^4=0.311 \pm 0.037$~\cite{SNO}. 
The experimental normalization uncertainty is assumed to be 10\% in JUNO, and 
Eq.~(\ref{eq:osc1}) is used as the oscillation formula.  
The true MC spectrum is generated assuming 
$|U_{e1}|^2+|U_{e2}|^2+|U_{e3}|^2 = 1$. The hypotheses when 
$|U_{e1}|^2+|U_{e2}|^2+|U_{e3}|^2 $ deviates from unity are then 
tested by fitting the MC data. The results are presented as 
$\Delta \chi^2 = \chi^2_{|U_{e1}|^2+|U_{e2}|^2+|U_{e3}|^2} -\chi^2_{unity}$ in Fig. 1. 
At 68\% C.L. ($\Delta \chi^2<1$), the combination of JUNO, Daya Bay, and SNO results 
would give about 4\% unitarity test. This unitarity test can be significantly
improved with a stronger constraint from solar neutrino experiments. For example, 
with a five times improved constraint of 
$|U_{e2}|^2\cdot(|U_{e1}|^2+|U_{e2}|^2) + |U_{e3}|^4$,  the unitarity test can 
be improved to about 1.2\% level, which will be sufficiently accurate to 
test the reactor anomaly~\footnote{\textcolor{black}{If this unitarity test is 
shown to be violated with future experimental data, beside the existence of 
sterile neutrino, the non-standard interaction in the sun or other new physics
could also be considered as the explanation.}}. 


\paragraph{\bf Other unitarity tests:}

The rest of equations in Eq.~(\ref{eq:uni1}) (for the $\mu$ and $\tau$ flavor 
neutrinos) are more difficult to test. First, 
the only oscillation to be precisely measured in the foreseeable future is 
the $\nu_{\mu}$ disappearance in the $\Delta m^2_{3x}$ oscillations. 
Second, due to the small difference between $\Delta m^2_{31}$ and $\Delta m^2_{32}$, 
one would need a third independent constraint, in analogy to the solar neutrino 
measurements, even if the $\nu_{\mu}$ disappearance of the $\Delta m^2_{21}$ 
oscillation is determined. Finally, the unitarity tests would also suffer from the 
uncertainties in experimental absolute normalization, which however could be 
improved with a future neutrino factory~\cite{nustorm}. 

Direct unitarity tests of  Eq.~(\ref{eq:uni2}) can be accomplished by  combining information 
from disappearance and appearance channels. For example, we can square both sides of 
$U_{e1}U^{*}_{\mu1} +U_{e2}U^{*}_{\mu2} +U_{e3}U^{*}_{\mu3} \stackrel{?}{=} 0$:
\begin{eqnarray}\label{eq:sq}
0 &\stackrel{?}{=}& |U_{e1}|^2|U_{\mu1}|^2 + |U_{e2}|^2|U_{\mu2}|^2 + |U_{e3}|^2|U_{\mu3}|^2 \nonumber \\
   &+& 2 \Re \left( U_{e1}U^{*}_{\mu1} U_{\mu 2} U^{*}_{e 2} \right) 
   + 2 \Re \left( U_{e1}U^{*}_{\mu1} U_{\mu 3} U^{*}_{e 3} \right) \nonumber \\
   &+& 2 \Re \left( U_{e2}U^{*}_{\mu2} U_{\mu 3} U^{*}_{e 3} \right).
\end{eqnarray}
In order to directly test the above equation, one would need to measure
$|U_{ei}|^2|_{i=1,2,3}$ as well as $|U_{\mu i}|^2|_{i=1,2,3}$ from disappearance channels.  
The latter three terms can be in principle accessed 
through the measurement of $\nu_{\mu}$ to $\nu_e$ appearance probability. 
However, the current (No$\nu$a and T2K) and the next generation (LBNE and Hyper-K) 
experiments will only focus on the $\Delta m^2_{3x}$ oscillations, which 
leaves the $\Delta m^2_{21}$ oscillations unconstrained. 

Eqs.~(\ref{eq:uni3}) and ~(\ref{eq:uni4}) can be tested by combining 
measurements of $\nu_{l}$ disappearance and $\nu_{l}$ to $\nu_{l'}$ appearance. 
For example, the constant term of the summation of $\nu_{\mu}$ disappearance, $\nu_{\mu}$ to $\nu_e$ appearance, 
and $\nu_{\mu}$ to $\nu_{\tau}$ appearance oscillation probabilities
in vacuum would be the $\sum_i (|U_{ei}|^2 + |U_{\mu i}|^2 + |U_{\tau i}|^2) 
\cdot |U_{\mu i}|^2$, which is related to the Eq.~(\ref{eq:uni3}). However, this measurement 
would rely on an accurate determination of the experimental absolute normalization factor. 
Eq.~(\ref{eq:uni4}) can be tested by searching for the absence of $L/E$ dependence
in the summation of oscillation probabilities from all three channels. However, 
in practice, these tests suffer from the matter effects, from the tiny difference between 
$\Delta m^2_{32}$ and $\Delta m^2_{31}$, as well as from the limited precision of 
$\nu_{\mu}$ to $\nu_{\tau}$ appearance channel. Therefore, it is actually more practical 
to perform an indirect unitarity test through measurements of the $\theta_{23}$
or the $\theta_{13}$
mixing angles, 
as discussed in the following. 

\paragraph{\bf Indirect unitarity tests:}
  
While the direct unitarity tests appear to be extremely difficult, 
the violation of unitarity may be also naturally indicated in the next generation 
experiments searching for sterile neutrinos (e.g. Ref.~\cite{nucifer,*sbl_karsten,*kamland_st,*dayabay_st,*broxeno_st,*TPC_st,*sage,*isodar_st,*oscsns,*LAR1_st}). A discovery of sterile 
neutrino could be established by unambiguously observing new oscillation patterns 
(different from the known $\Delta m^2_{12}$ and $\Delta m^2_{3x}$ oscillations).  
\textcolor{black}{Most currently proposed searches are focusing on a limited range 
(e.g. $\sim$1 eV) of the sterile neutrino masses motivated by the 
various anomalies in the data. }
On the other hand, if the sterile neutrinos or other new physics existed, 
the current measured mixing angles in the PMNS matrix would be effective angles,
as discussed above, 
whose values would be process dependent. This point has been raised 
for example in Refs.~\cite{st1,st2} before. 
In such kind of tests, the 
``proof by contradiction'' principle is utilized. First, the mixing angles 
are extracted from the data by assuming unitarity. If the same mixing angle 
measured by two different processes are inconsistent, the unitarity 
is then shown to be violated. Otherwise, the phase space of new physics will 
be constrained. Here, the word "indirect" comes from the fact that components in
Eqs.~\eqref{eq:uni1}-\eqref{eq:uni4} are not measured. 

There are currently three possibilities for such indirect unitarity tests: 
$\theta_{23}$, $\theta_{12}$, and $\theta_{13}$. The $\theta_{23}$ indirect test 
can be achieved by comparing the $\nu_{\mu}$ disappearance and $\nu_{\mu}$ to $\nu_{\tau}$ 
appearance. The precision will be limited by the $\nu_{\tau}$ appearance channel. 
For $\theta_{12}$, the indirect test between solar neutrino and medium-baseline 
reactor experiment is not necessary, as the direct test can be carried out as illustrated above.
 
Therefore, the most promising candidate for such indirect unitarity test is $\theta_{13}$,
which can be measured with the $\nu_{\mu}$ to $\nu_e$ 
appearance (T2K, No$\nu$a, LBNE, and Hyper-K) with accelerator neutrinos, 
as well as the $\bar{\nu}_e$ disappearance with reactor neutrinos 
(Daya Bay, RENO, and Double Chooz). For example, one can test the 
hypothesis of the 6\% reactor anomaly due to the fourth generation sterile neutrino. 
From Ref.~\cite{st2}, the $\nu_\mu \rightarrow \nu_e$ appearance probability 
in vacuum will be altered by the additional sterile (fourth) neutrino  as:
\begin{eqnarray}\label{eq:ap1}
P &=& 4|U_{\mu3}|^2|U_{e3}|^2\sin^2\left ( \frac{\Delta m^2_{31} L}{4E}\right)    \nonumber\\
 &+& 4|U_{\mu2}|^2|U_{e2}|^2\sin^2\left ( \frac{\Delta m^2_{21} L}{4E}\right) \nonumber\\
 &+& 8|U_{\mu3}| |U_{e3}| |U_{\mu2}| |U_{e2}|  \nonumber\\
&\times& \sin  \left ( \frac{\Delta m^2_{31} L}{4E}\right) \sin \left ( \frac{\Delta m^2_{21} L}{4E}\right)
\cos \left ( \frac{\Delta m^2_{32} L}{4E} -\delta_3 \right) \nonumber \\
&+&4|U_{\mu3}||U_{e3}||\beta''| \sin  \left ( \frac{\Delta m^2_{31} L}{4E}\right) \sin  \left ( \frac{\Delta m^2_{31} L}{4E} - \delta_1\right) \nonumber \\
&+&4|U_{\mu2}||U_{e2}||\beta''| \sin  \left ( \frac{\Delta m^2_{21} L}{4E}\right) \sin  \left ( \frac{\Delta m^2_{21} L}{4E} - \delta_2\right) \nonumber \\
&+& 2|U_{\mu4}|^2|U_{e4}|^2,
\end{eqnarray}
where $\beta''= U^*_{\mu4} U_{e4}$, $\delta_1 = -\arg(U_{\mu3}U^*_{e3}\beta'')$, 
$\delta_2 = -\arg(U_{\mu2}U^*_{e2}\beta'')$, and $\delta_3 = \arg(U^*_{\mu3} U_{e3} U_{\mu 2} U^*_{e2})$. 
The only approximation made here is to average terms containing large sterile mass squared differences.
If we neglect the $\Delta m^2_{21}$ oscillations and assume that $\delta_1 = 0$, the change in the 
effective $\sin^22\theta_{13}$ from the long-baseline $\nu_e$ appearance experiment can be estimated as 
$\frac{\Delta \sin^2 2\theta_{13}}{ \sin^2 2\theta_{13}} = \frac{|U_{\mu4}||U_{e4}|}{ |U_{\mu3}| |U_{e3}|}$.
If we assume $U^*_{\mu4} U_{e4}=0.04$, which satisfies the 90\% C.L. constrained from the latest ICARUS 
experiment~\cite{icarus} ($2|U_{\mu4}|^2|U_{e4}|^2 < 3.4\times 10^{-3}$), the 
effective $\sin^22\theta^{eff}_{13}$ in the appearance channel could be higher than 
the true one by as much as 40\%. On the other hand, given the 6\% reactor anomaly, 
the effective $\sin^22\theta^{eff}_{13}$ obtained through the reactor antineutrino disappearance 
experiments could be only about a few percent higher than the true one (e.g. $\sin^22\theta^{eff}_{13} = 
\frac{\sin^22\theta_{13}}{0.94}$ with $\sin^22\theta_{13} := 4|U_{e3}|^2\cdot \left(|U_{e1}|^2+|U_{e2}|^2 \right)$)~\footnote{\textcolor{black}{In our estimation, since both $U_{e4}$ and $U_{\mu4}$
are small, we have assumed that values of $|U_{e3}|^2|U_{\mu3}|^2$ and $4|U_{e3}|^2\cdot \left(|U_{e1}|^2+|U_{e2}|^2 \right)$ stay the same when we change from 3-neutrino model to 3-neutrino + 1-sterile 
neutrino model. This is not exactly true, as the components of the 3x3 PMNS matrix will change with 
non-zero $U_{e4}$ and $U_{\mu4}$. Additional small corrections should be applied. Nevertheless, 
these will not alter our conclusion.}}.
In comparison, the projected precision of the Daya Bay experiment~\cite{dayabay_proceeding}, 
LBNE10~\footnote{LBNE10 represents the phase I of the LBNE program. LBNE10 contains a 10 kt 
liquid argon time projection chamber. The running time includes 5 years neutrino and 5 years 
antineutrino running with a 708 kW beam. }, and full LBNE is about $<$4\%, $\sim$10\%, and $<$5\%, respectively.  
Therefore, by comparing the measured $\sin^22\theta_{13}$ value from the reactor experiments
to that measured in accelerator experiments, one would rule out the specific
hypothesis described above, given the unitarity is truly conserved. 

The recent T2K $\nu_e$ appearance results~\cite{T2K_ap1} favors a larger value of $\sin^22\theta_{13}$
than that from the reactor $\bar{\nu}_e$ disappearance results. The statistical 
significance is at about  2$\sigma$ level, whose actual value would depend on the assumption of the mass 
hierarchy and the value of CP phase $\delta$. Such a difference at present is consistent 
with an explanation of a statistical fluctuation. If the difference persists 
with increased statistics, the hypothesis of existence of new physics would be favored. 
Otherwise, the phase space of new physics can be strongly constrained. Furthermore, as shown in 
Eq.~(\ref{eq:ap1}), the existence of the fourth generation of sterile neutrino will likely not 
only change the effective mixing angle, but will also introduce additional spectrum distortion through 
non-zero $\delta_1$ or $\delta_2$ phases. Therefore, the wide band beam of LBNE together with its 
high statistics measurement of disappearance/appearance spectra would provide stringent tests 
for new physics. 

There is actually another group of indirect unitarity tests. For example, 
one can see that Eq.~(\ref{eq:sq}) 
is the same one as $P(\nu_{\mu}\rightarrow \nu_e)$ at $L=0$. Therefore, the search for appearance of $\nu_e$
with low backgrounds at very short baseline would effectively test unitarity. Such an experiment (e.g. 
ICARUS~\cite{icarus}) is indeed very powerful in constraining the phase space of sterile 
neutrino models, which are motivated by the LSND~\cite{LSND}, MiniBooNe~\cite{miniboone}, and reactor~\cite{anom} anomalies.

\paragraph{\bf Conclusions:}

In this paper we illustrate the direct and indirect unitarity tests of 
the PMNS matrix with a few simple examples. In order to calculate the sensitivity 
of direct unitarity test of the first row $|U_{e1}|^2+|U_{e2}|^2+|U_{e3}|^2 \stackrel{?}{=} 1$, 
we approximate SNO results 
as a measurement of $|U_{e2}|^2\cdot(|U_{e1}|^2+|U_{e2}|^2) + |U_{e3}|^4$. 
A critical assessment of this formula can be found in Ref.~\cite{baha}.
We also neglect the matter effects in the long baseline $\nu_e$ appearance 
measurement in illustrating the power of indirect unitarity tests with $\theta_{13}$.

Although direct 
unitarity tests appear to be extremely challenging given limited experimentally
available oscillation channels, we show that the combination of the
medium-baseline reactor experiment, short-baseline reactor experiments, and the SNO 
solar results will make it possible to perform the first direct and model independent
unitarity test of the PMNS matrix. 
At 68\% C.L., the combination of JUNO, Daya Bay, and SNO will test 
$|U_{e1}|^2+|U_{e2}|^2+|U_{e3}|^2=1$  at a 4\% level. This level of accuracy can be substantially 
reduced with an improved constraint from solar neutrino measurements. 
In addition, by comparing the $\sin^22\theta_{13}$ 
values measured by the current generation reactor neutrino experiment vs. current/next generation 
accelerator neutrino experiments, one can perform an indirect unitarity test, which would put 
strong constraints on the possible new physics (e.g. sterile neutrino, non-standard 
interaction etc.) beyond the three-neutrino model. Such constraints will be further
enhanced by the precision measurement of disappearance/appearance spectra with a wide 
band beam.

\paragraph{\bf Acknowledgments:} We would like to thank B. Viren and R. D. McKeown for fruitful discussions. 
This work was supported in part by the National Science
Foundation, and the Department of Energy under contract DE-AC02-98CH10886.

\bibliographystyle{unsrt}
\bibliography{unitarity}{}

\end{document}